\documentclass[twocolumn]{jpsj2}

\def\runtitle{Dynamical Susceptibility in KDP-type Crysals}
\def\runauthor{Shun-ichi \textsc{Yoshida} , Norihiro \textsc{Ihara} , Koh \textsc{Wada}}

\title{%
Dynamical Susceptibility in KDP-type Crysals above and below
$T_{\rm c}$ II
}
\author{%
Shun-ichi \textsc{Yoshida}\thanks{E-mail:
shun1@statphys.sci.hokudai.ac.jp}, 
Norihiro \textsc{Ihara}
 and Koh
\textsc{Wada}
}

\inst{%
Division of Physics, Graduate School of Science,
 Hokkaido University, Sapporo 060-0810
}

\recdate{\today}

\abst{%
The path probability method (PPM) in the tetrahedron-cactus
 approximation is applied to the Slater-Takagi model with dipole-dipole
  interaction for $\rm KH_{2}PO_{4}$-type hydrogen-bonded ferroelectric
   crystals in order to derive a small dip structure in the real part of 
    dynamical susceptibility observed at the transition temperature
     $T_{\rm c}$.
The dip structure can be ascribed to finite relaxation times of electric
 dipole moments responsible for the first order transition with contrast
  to the critical slowing down in the second order transition.
The light scattering intensity which is related to the imaginary part of
 dynamical susceptibility is also calculated above and below the
  transition temperature and the obtained central peak structure is
   consistent with polarization fluctuation modes in Raman scattering
    experiments.
}

\kword{%
KDP(KH$_2$PO$_4$),
Slater-Takagi model,
PPM(path probability method),
dynamical susceptibility,
light scattering intensity
}

\begin{document}
\sloppy
\maketitle

\section{Introduction}
In the previous paper\cite{Wada}(this article is referred to as I
 henceforth), 
we have succeeded in calculating the dynamical
 susceptibility for the Slater-Takagi model\cite{Matsubara} above and
  below the transition temperature $T_{\rm c}$ by making use of the
   analytical solution for spontaneous polarization in the
    tetrahedron-cactus approximation of the cluster variation
     method(CVM)\cite{Ishibashi}.
This approximation is the basic one in the CVM for taking into account
 the ice-rule for protons of KDP-type hydrogen bonded ferroelectrics.
However, since the Slater-Takagi model always yields the second order
 phase transition, the dynamical susceptibility at $T=T_{\rm c}$ almost
  vanishes as a result of critical slowing down of the most dominant mode
   governing the transition.
This result is different from the experimental data of KDP
 (KH$_2$PO$_4$) which show only a small dip at $T=T_{\rm c}$ in the
  real part of dynamical susceptibility versus temperature
   graph\cite{Koz}.

It is believed that KDP undergoes the first order transition close to
 the second order. 
Senko\cite{Senko} phenomenologically introduced the dipole-dipole
 interaction to the Slater-Takagi model and Silsbee {\it et al.}\cite{Silsbee}
  revealed that the phase transition of this model with dipole-dipole
   interaction changes from the second order to the first order
    transition according to increase of the strength of dipole-dipole
     interaction. 
Based on this model Matsumoto and Ogura\cite{Matu} have calculated the
 light scattering intensity related to the imaginary part of the
  dynamical susceptibility in the tetrahedron-cactus approximation.
However, their calculation seems to be limited to the paraelectric
 phase. 
In the present manuscript we would like to show the small dip structure
 at the transition temperature in the experimental data of the real part
  of dynamical susceptibility versus temperature for KDP-type crystal.
We will also calculate the light scattering intensity below the
 transition temperature as well as above it.
For that purpose we apply the tetrahedron cactus approximation of the
 PPM (Path Probability Method)\cite{Kiku1} to the Slater-Takagi model
  with phenomenologically induced dipole-dipole interaction since we
   expect that the most dominant mode governing the transition brings
    about a finite relaxation time at the transition temperature.

\section{Formulation}
\begin{figure}[tbp]
\begin{center}
\includegraphics[width=8cm]{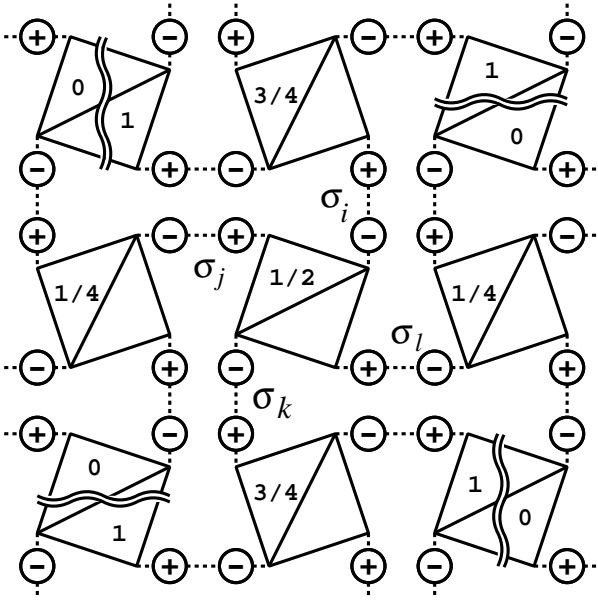}
\caption{
The projection of atomic arrangement of KDP-type crystals on (001)
 plane.
The variables  $\sigma_i, \sigma_j, \sigma_k, \sigma_l$ show the four
 different pseudo-spins for protons around a PO$_4$ tetrahedron.
The number described in a PO$_4$ tetrahedron represents a relative
 height of a PO$_4$ tetrahedron.
}
\label{2D}
\end{center}
\end{figure}

There are $N$ PO$_4$ tetrahedra and $2N$ protons in KDP-type crystals as
shown in Fig. \ref{2D}.
The pseudo-spin Ising Hamiltonian $\mathcal{H}$ for a configuration of $2N$
 protons has a form
\begin{equation}
\begin{split}
 \mathcal{H}
 &=
 \sum_{\langle ijkl\rangle}
 \left(
  \mathcal{H}_0(\sigma_i,\sigma_j,\sigma_k,\sigma_l)
  - \frac{\mu_{\rm d}E}{2}
    (\sigma_{i}+\sigma_{j}+\sigma_{k}+\sigma_{l})\right)\\
 &\quad+{\cal H}_{\rm dipole},
\end{split}
\end{equation}
with
\begin{multline}
\mathcal{H}_0(\sigma_i,\sigma_j,\sigma_k,\sigma_l)\\
= -V_2(\sigma_i\sigma_j+\sigma_j\sigma_k+\sigma_k\sigma_l+\sigma_l\sigma_i)\\
 -V_4\sigma_i\sigma_j\sigma_k\sigma_l-V_5(\sigma_i\sigma_k+\sigma_j\sigma_l)+C
\label{H1},
\end{multline}
and 
\begin{equation}
\begin{split}
 \mathcal{H}_{\rm dipole}
 &=
 -\sum_{\langle ijkl\rangle}
 \lambda{\mu_{\rm d}}^2 \,m\, 
 (\sigma_{i}+\sigma_{j}+\sigma_{k}+\sigma_{l})\\
 &\quad +2N\lambda{\mu_{\rm d}}^2 m^2\,,
\end{split}
\label{findip}
\end{equation}
where the sum $\langle ijkl\rangle$ runs over four protons $i,j,k,l$
 around each $\rm PO_{4}$ tetrahedron in the crystal,
$\mu_{\rm d}$ is the magnitude of an electric dipole moment associated
 with a complex $\rm K$-$\rm H_{2}PO_{4}$,
$E$ is an external electric field, $\lambda$ is a parameter
 representing long range dipolar sum in the phenomenologically
  introduced dipole-dipole interaction $\mathcal{H}_{\rm dipole}$ and
   $m$ is the electric polarization defined by thermal average over
    $\mathcal{H}$ as
\begin{equation}
m
 = \langle\sigma_i\rangle
 = \langle\sigma_j\rangle
 = \langle\sigma_k\rangle
 = \langle\sigma_l\rangle \,,
\end{equation}
and $2N\lambda{\mu_{\rm d}}^2 m^2$ in $\mathcal{H}_{\rm dipole}$ is taken 
care of over-counting due to molecular field.

\begin{figure}[t]
\begin{center}
\includegraphics[width=8cm]{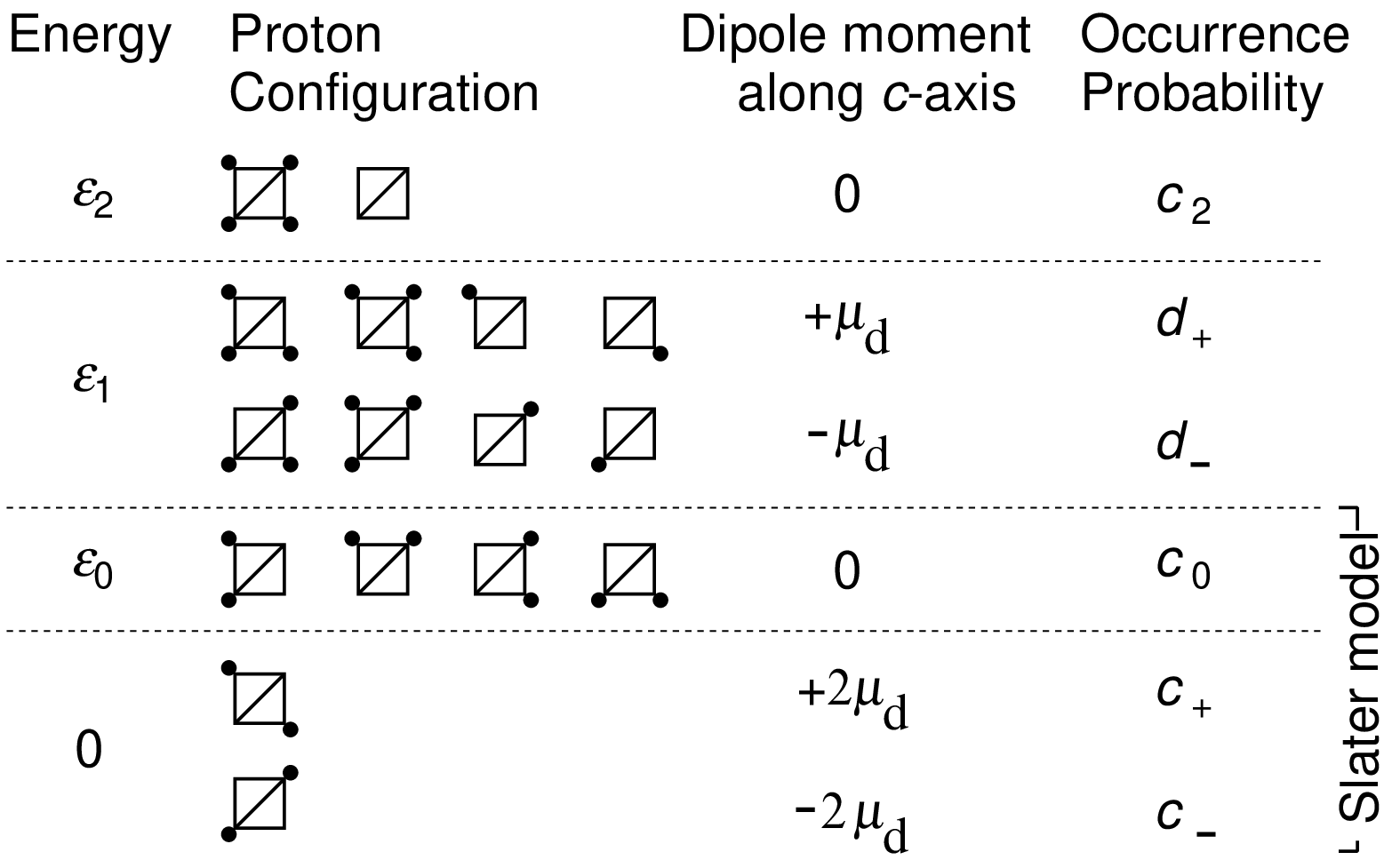}
\caption{Configuration energy
 ($\varepsilon_{2}>\varepsilon_{1}>\varepsilon_{0}>0$), alloted dipole
  moment along $c$-axis and occurrence probability of proton
   configuration around $\rm PO_{4}$ tetrahedron.}
\label{ST}
\end{center}
\end{figure}

As is seen in Fig. \ref{2D},
we use a convention that when the $i$-th proton is located on the closer
 site to an O atom at the bottom (top) of the $\rm PO_{4}$ tetrahedron
  along the easy $c$-axis, the $i$-th proton takes $\sigma_{i}=+1(-1)$.
The energy parameters $V_{2}, V_{4}, V_{5}$ and $C$ are connected to the 
energy parameters in the Slater-Takagi model shown in Fig. \ref{ST} as
\begin{equation}
\begin{split}
& V_{2}=\varepsilon_{2}/8 \,,\quad
  V_{4}=-\varepsilon_{0}/4+\varepsilon_{1}/2-\varepsilon_{2}/8 \,,\\
& V_{5}=\varepsilon_{0}/4-\varepsilon_{2}/8 \,,\quad
  C=\varepsilon_{0}/4+\varepsilon_{1}/2+\varepsilon_{2}/8 \,.
\end{split}
\end{equation}

As shown in I,
 we now apply the path probability method (PPM)\cite{Kiku1} in the
  tetrahedron-cactus approximation to the present system in order to
   find the time evolution equation of the system with  various proton
    configurations.
After some manipulations of the PPM we can make a generating function
 from which a set of time evolution equations is derived through
  differentiation of an interaction parameter set
   $\mib{L}\equiv(L_{1},\,L_{2},\, L_{3},\, L_{4},\, L_{5})$. 
The generating function is given by 
\begin{equation}
\begin{split}
&\mathcal{G}(\mib{L})
 = \frac{1}{\tau_{0}}\mathop{\rm Tr}_i p_{1}(\sigma_i,t)e^{-2L_{1}\sigma_i}\\
&\qquad
 \times
 \left[
 \mathop{\rm Tr}_{jkl}
 \frac{
  p_4(\sigma_i, \sigma_j, \sigma_k, \sigma_l,t)\ e^{-\frac{\beta}{2}{\it\Delta}_i \mathcal{H}_0(\sigma_i, \sigma_j, \sigma_k, \sigma_l)}
 }{
 p_1(\sigma_i,t)
 }\right]^2\,,\\[5mm]
&\mathit{\Delta}_{i}\mathcal{H}_{0}(\sigma_i, \sigma_j, \sigma_k, \sigma_l)\\
&\qquad =\mathcal{H}_{0}(-\sigma_i, \sigma_j, \sigma_k, \sigma_l)-{\cal H}_{0}(\sigma_i, \sigma_j, \sigma_k, \sigma_l), \\
 &L_{1}
 = \frac{\,1\,}{\,2\,}
   \bigl(
   \beta\mu_{\rm d}E+2\beta\lambda\mu^{2}_{\rm d}m(t)
   \bigr)\,,\\
 &L_2 = \beta V_2\,,\quad
  L_3 = \beta V_3\,,\quad
  L_4 = \beta V_4\,,\quad
  L_5 = \beta V_5\,,
\end{split}
\end{equation}
where $\mathop{\rm Tr}_{i}$ and $\mathop{\rm Tr}_{jkl}$ denotes a trace
 operation $\sum_{\sigma_i=\pm 1}$ and $\sum_{\sigma_j, \sigma_k,
  \sigma_l=\pm 1}$, respectively,
$\beta=1/k_{\rm B}T$ is the inverse temperature with Boltzmann's
 constant $k_{\rm B}$,
${\it\Delta}_i\mathcal{H}_{0}(\sigma_i,\sigma_j,\sigma_k,\sigma_l)$
 defines an energy change under transfer of $i$-th proton between the
  double well potential on a hydrogen bond and $\tau_{0}$ is its
   microscopic relaxation time of an isolated proton.
When $\lambda=0$, the generating function $\mathcal{G}(\mib{L})$
 coincides with that of I.
Thus, the effect of dipole-dipole interaction  modifies only the
 external electric field $E$ by $E+2\lambda\mu_{\rm d}m(t)$.
Further, state variables $p_1(\sigma_i,t)$ and
 $p_4(\sigma_i,\sigma_j,\sigma_k,\sigma_l,t)$ represent, respectively, the
  probability of finding the site $\sigma_i$ of a proton in the $i$-th
   hydrogen bond at time $t$ and the probability of finding the sites
    $\sigma_i, \sigma_j, \sigma_k, \sigma_l$ of protons $i, j, k, l$
     around a PO$_4$ tetrahedron at time $t$.
The state variables $p_1$ and $p_4$ are given by
\begin{equation}
\begin{split}
 &p_{1}(\sigma_i,t)=\frac{\,1\,}{\,2\,}\bigl(1+m(t)\sigma_i\bigr),\\
 &p_{4}(\sigma_i,\sigma_j,\sigma_k,\sigma_l,t)\\
&=
\frac{\,1\,}{\,2^4\,}
 \biggl(
 1
 +\frac{m_1(t)}{4}(\sigma_i+\sigma_j+\sigma_k+\sigma_l)\\
&\qquad\qquad
 +\frac{m_{2}(t)}{4}
  (\sigma_i\sigma_j+\sigma_j\sigma_k+\sigma_k\sigma_l+\sigma_l\sigma_i)\\
&\qquad\qquad
 +\frac{m_{3}(t)}{4}
  (\sigma_i\sigma_j\sigma_k+\sigma_j\sigma_k\sigma_l
   +\sigma_k\sigma_l\sigma_i+\sigma_l\sigma_i\sigma_j)\\
&\qquad\qquad
 +m_{4}(t)\sigma_i\sigma_j\sigma_k\sigma_l
 +\frac{m_{5}(t)}{2}(\sigma_i\sigma_k+\sigma_j\sigma_l)
 \biggr).
\end{split}
\end{equation}
Here let us introduce the Ising spin space vector $\mib{\sigma}$ around
a PO$_4$ tetrahedron as
\begin{eqnarray}
\mib{\sigma} = 
\begin{pmatrix}
 \sigma_i+\sigma_j+\sigma_k+\sigma_l\\
 \sigma_i\sigma_j+\sigma_j\sigma_k+\sigma_k\sigma_l+\sigma_l\sigma_i\\
 \sigma_i\sigma_j\sigma_k+\sigma_j\sigma_k\sigma_l
  +\sigma_k\sigma_l\sigma_i+\sigma_l\sigma_i\sigma_j\\
 \sigma_i\sigma_j\sigma_k\sigma_l\\
 \sigma_i\sigma_k+ \sigma_j\sigma_l               
\end{pmatrix}
\end{eqnarray}
Then, in the tetrahedron-cactus approximation of the PPM, the
 homogeneous state of the present system at time $t$ is found to be
  described by the five order parameters defined by
\begin{eqnarray}
\begin{pmatrix}
 m_1\\
 m_2\\
 m_3\\
 m_4\\
 m_5
\end{pmatrix}
 = 
\begin{pmatrix}
 \langle \sigma_i+\sigma_j+\sigma_k+\sigma_l \rangle\\
 \langle \sigma_i\sigma_j+\sigma_j\sigma_k
  +\sigma_k\sigma_l+\sigma_l\sigma_i\rangle\\
 \langle \sigma_i\sigma_j\sigma_k+\sigma_j\sigma_k\sigma_l
  +\sigma_k\sigma_l\sigma_i+\sigma_l\sigma_i\sigma_j \rangle\\
 \langle \sigma_i\sigma_j\sigma_k\sigma_l \rangle\\ 
 \langle \sigma_i\sigma_k+ \sigma_j\sigma_l \rangle               
\end{pmatrix}
\end{eqnarray}
where each order parameter represents the correlation of protons $i, j, k, l$ 
 around a $\rm PO_{4}$ tetrahedron at time $t$,
  $\langle\cdots\rangle_{t}$ is a thermal average at time $t$ and
   $m(t)=m_1(t)/4$ is the long range order parameter due to electric
    polarization per proton. 
Then a set of kinetic equations for five order parameters is given in a
 convenient form as 
\begin{equation}
\frac{{\rm d}m_{i}(t)}{{\rm d}t}
=4 \lim_{L_{3}\to 0}
 \frac{\partial \mathcal{G}(\mib{L})}{\partial L_{i}}
\qquad (i=1\sim 5).
\label{eqmo}  
\end{equation}
Here it should be noted that in order to write the above expression an
 extra interaction term is virtually added to Hamiltonian (\ref{H1}) for
  technical convenience as
\begin{equation}
\begin{split}
 &\mathcal{H}_{0}(\sigma_i,\sigma_j,\sigma_k,\sigma_l)\\
 &\qquad -V_{3}(\sigma_i\sigma_j\sigma_k+\sigma_j\sigma_k\sigma_l+\sigma_k\sigma_l\sigma_i+\sigma_l\sigma_i\sigma_j)\\
 &\to \mathcal{H}_{0}(\sigma_i,\sigma_j,\sigma_k,\sigma_l),
\end{split}
\end{equation}
and $V_{3}$ is, however, put to zero just after differentiation with
 respect to $L_{3}$ in eq.(\ref{eqmo}).

\section{Static Properties}
As shown in I, in order to obtain dynamical susceptibility
 $\chi(\omega)$ as a linear response to an external field $E$,
  equilibrium values of order parameters are required at each
   temperature.
Since the equilibrium state is more easily obtained from the cluster
 variation method (CVM)\cite{Kiku2} rather than from the stationary
  solution of the time evolution equations(\ref{eqmo}),
we apply the tetrahedron-cactus approximation of the CVM to the present
 system\cite{Ihara}.
The variational free energy $G$ is obtained by
\begin{eqnarray}
G=U-TS,
\end{eqnarray}
where the internal energy $U$ is given by
\begin{equation}
\begin{split}
&\frac{U}{N}
 = -V_{2}m_{2}-V_{4}m_{4}-V_{5}m_{5}
   -{2\mu_{\rm d}E}m_{1}-2\lambda\mu_{\rm d}^2 m^2_{1},
\end{split}
\end{equation}
and the entropy $S$ is given by
\begin{equation}
\begin{split}
 &\frac{S}{Nk_{\rm B}}
 =
 \mathop{\rm Tr}_{i}
 \bigl[
  p_1(\sigma_i)\ln p_1(\sigma_i)
 \bigr]\\
 &\qquad\qquad
 -\mathop{\rm Tr}_{ijkl}
 \bigl[
  p_{4}(\sigma_i,\sigma_j,\sigma_k,\sigma_l)
  \ln p_{4}(\sigma_i,\sigma_j,\sigma_k,\sigma_l)
 \bigr].
\end{split}
\end{equation}
Here $c_+, c_-, c_0, d_+,  d_-$ and $c_2$ shown in
 Fig. \ref{ST} are defined through
  $p_{4}(\sigma_i,\sigma_j,\sigma_k,\sigma_l)$ as
\begin{equation}
\begin{split}
 c_+&= \frac{1}{2^4}(1+m_1+m_2+m_3+m_4+m_5),\\
 c_-&= \frac{1}{2^4}(1-m_1+m_2-m_3+m_4+m_5),\\
 c_0&= \frac{1}{2^4}(1+m_4-m_5),\\
 d_+&= \frac{1}{2^4}(1+m_1/2-m_3/2-m_4),\\
 d_-&= \frac{1}{2^4}(1-m_1/2+m_3/2-m_4),\\
 c_2&= \frac{1}{2^4}(1-m_2+m_4+m_5),
\end{split}
\end{equation}
with a normalization condition
\begin{eqnarray}
c_+ + c_- + 4c_0 + 4 (d_+ + d_-)+2 c_2=1. \label{normal}
\end{eqnarray}
Since it is sometimes more convenient to use 
 $c_+, c_-, c_0, d_+, d_-$ and $c_2$ instead of $m_{1}\sim m_{5}$
 especially in thermostatic discussions, 
  we rewrite the variational free energy in terms of
   $c_+, c_-, c_0, d_+, d_-$ and $c_2$ as
\begin{equation}
\begin{split}
 \frac{G}{N}
 &=
 4\varepsilon_{0}c_{0}
 +4\varepsilon_{1}(d_{+}+d_{-})
 +2\varepsilon_{2}c_{2}\\
 &\quad
 -2{\lambda\mu_{\rm d}}^2(p_{+}-p_{-})^2
 -2\mu_{\rm d}E(p_{+}-p_{-})\\
 &\quad
 -k_{\rm B}T(2p_{+}\ln p_{+}+2p_{-}\ln p_{-}
 -c_{+}\ln c_{+}-c_{-}\ln c_{-}\\
 &\quad
  -4c_{0}\ln c_{0}
  -4d_{+}\ln d_{+}-4d_{-}\ln d_{-}
  -2c_{2}\ln c_{2})\\
 &\quad
 +s\bigl(
  1-\bigl(c_{+}+c_{-}+4c_{0}+4(d_{+}+d_{-})+2c_{2}\bigr)
  \bigr),\label{G}
\end{split}
\end{equation}
with
\begin{equation}
\begin{split}
 p_{1}(+1)&\equiv p_+ = c_+  + 2c_0 +3 d_+ + d_- +c_2,\\
 p_{1}(-1)&\equiv p_- = c_- + 2c_0 + d_+ +3 d_- +c_2,
\end{split}
\end{equation}
where $s$ is the Lagrange multiplier to make all the state variables
 $c_+, c_-, c_0, d_+, d_-$ and $c_2$ independent.
With $\lambda=0$ this expression of the free energy reduces to that of
 Ishibashi.\cite{Ishibashi}
Under a uniform electric field the thermal equilibrium state is obtained
 from the minimum condition of the free energy:
$
 {\partial G}/{\partial c_{+}}
 = {\partial G}/{\partial c_{-}}
 = {\partial G}/{\partial c_{0}}
 = {\partial G}/{\partial d_{+}}
 = {\partial G}/{\partial d_{-}}
 = {\partial G}/{\partial c_{2}} = 0$
as 
\begin{equation}
\begin{split}
& c_0 = \frac{\eta_0}{4\eta_0+2\eta_2+((hA)^2+(hA)^{-2})+4\eta_1(hA+(hA)^{-1})}\,,\\
& c_+ = (hA)^2 \left(\frac{c_0}{\eta_0}\right),\,\,
  c_- = (hA)^{-2} \left(\frac{c_0}{\eta_0}\right),\,\,
  c_2 = \eta_2 \left(\frac{c_0}{\eta_0}\right),\\
& d_+ = (hA) \eta_1 \left(\frac{c_0}{\eta_0}\right),\,\,
  d_- = (hA)^{-1} \eta_1 \left(\frac{c_0}{\eta_0}\right)
\end{split}
\label{mkai}
\end{equation}
where 
\begin{equation}
\begin{split}
& \eta_i =\exp(-\beta\varepsilon_i)\,, \quad (i=0,1,2)\\
& A = \sqrt{\frac{1+m}{1-m}}
      \exp(2Dm)\,,\qquad(D=\beta\lambda{\mu_{\rm d}}^2)\\
& h= \exp(\beta\mu_{\rm d}E)\,.
\end{split}
\end{equation}
Further, the polarization $m$ is determined by the relation
\begin{eqnarray}
m = p_+ - p_- = c_+ - c_- + 2(d_+ - d_-)
\label{polarization}
\end{eqnarray}

\begin{figure}[btp]
\begin{center}
\includegraphics[width=8cm]{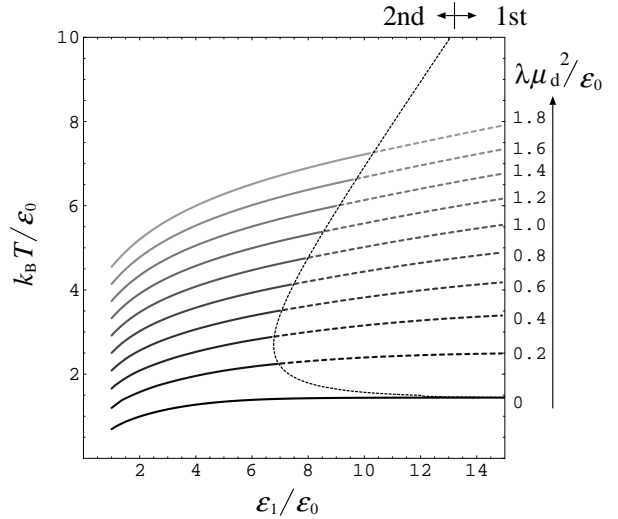}
 \caption{ The phase diagram of
 the Slater-Takagi model with dipole-dipole interaction. The solid line
 stands for the second order phase transition line and the dotted one
 stands for the first order phase transition.  } \label{phase}
\end{center}
\end{figure}

In order to investigate properties of the phase transition we make the
 Landau-type variational free energy $G(m,E,T)$ as a function of the
  electric polarization $m$ by substituting (\ref{mkai}) into
   eq.(\ref{G}):
\begin{equation}
\begin{split}
 &\frac{G(m,E,T)}{Nk_{\rm B}T}\\
 &=2Dm^2-\ln{\frac{1-m^2}{4}}\\
 &\qquad -\ln \left((A^2+A^{-2})+4\eta_0+4\eta_1(A+A^{-1})+2\eta_2\right)
 \label{Landau}
\end{split}
\end{equation}
It is noteworthy that $G(m,E,T)$ has a property
 $\partial G(m,E,T)/\partial m =0$ at the thermal equilibrium state.
Now, we expand $G(m,E,T)$ in terms of order parameter $m$ up to the 4-th
 order:
\begin{multline}
\frac{G(m,E,T)}{Nk_{\rm B}T}
= -\ln{\frac{\Gamma}{2}}+\frac{1}{2}A_2(T)m^2+\frac{1}{4}A_4(T)m^4+\cdots\\
- \frac{4\mu_d(1+2D)(1+\eta_1)}{\Gamma k_BT}mE
\end{multline}
where $\Gamma, A_2(T)$ and $A_4(T)$ are given by 
\begin{equation}
\begin{split}
& \Gamma
 = 1+2\eta_0+4\eta_1+\eta_2\,,\\
& A_2(T)
 = 2(1+2D)\left(1-\frac{2(1+2D)(1+\eta_1)}{\Gamma}\right)\,,\\
& A_4(T)
 = \frac{A_2(T)}{1-2D}+\frac{2(1+2D)}{\Gamma^2}\Bigl(4(1+2D)^3(1+\eta_1)^2\\
&\qquad\qquad -\frac{\Gamma}{3}\bigl(2(1+\eta_1)+(4+\eta_1)(1+2D)^3\bigr)\Bigr)
\end{split}
\label{state}
\end{equation}
From the view point of Landau's phase transition theory the order of the
 phase transition is classified as follows.
(i) The phase transition undergoes the second order transition at
 $T_0$ if $A_2(T_0)=0$ and $A_4(T_0)>0$.
(ii) The phase transition undergoes the first order phase transition at
 $T_{\rm c}(>T_0)$ if $A_2(T_0)=0$ and $A_4(T_0)<0$.
The boundary between the first and the second order is called the
 tricritical point $T_{\rm t}(=T_0)$ when $A_2(T_0)=0$ and
  $A_4(T_0)=0$.
Based on eq.(\ref{state}) we can get the phase diagram in
 Fig. \ref{phase}.
Since an increase of the phenomenologically introduced dipole-dipole
 interaction strengths the mechanical ferroelectric order effect, 
the ferroelectric transition becomes possible before the system comes to
 an instability point when the temperature is decreased.
This is the reason why the dipole-dipole interaction induces the first
 order transition instead of the second order transition.
As is shown in Fig.\ref{phase}, with $\varepsilon_{1}/\varepsilon_{0}$
 being decreased for each $\lambda$, the transition temperature is
  decreased because more easily thermal excitation of protons favors the
   paraelectric phase.
When $A_{2}(T_0)=0$ for $\lambda\mu_d^2/\varepsilon_0=0$ (Slater-Takagi
 model), $A_{4}(T_0)=\eta_1/(1+\eta_1)$ is positive definite.
That is, the Slater-Takagi model always undergoes the second order phase
 transition.
Especially, in the Slater model ($\lambda=0, \varepsilon_{0}>0,
 \varepsilon_2> \varepsilon_{1}\to\infty$) the phase transition takes
  place at the tricritical point.
Therefore, it is sometimes said that the transition derived by the
 Slater theory has the nature of both the first and the second order
  phase transitions.
On the other hand, as is seen in eq.(\ref{state}) the first order phase
 transition is caused by the finite value
  $\lambda{\mu_{\rm d}}^2/\varepsilon_0>0$.
As $\lambda{\mu_{\rm d}}^2/\varepsilon_{0}$ is increased, the first order
 transition region becomes larger and the transition temperature
  $T_{\rm c}$ is also increased.
However, when the dipolar term becomes dominant, the second order
 transition region again widens because of the phenomenologically
  induced dipolar interaction.
 
In order to fix the values of unkown parameters we utilize the
 experimental data on the temperature dependence of spontaneous
  polarization\cite{Reese} and choose the parameters such as
   $\varepsilon_{1}/\varepsilon_{0}=9.48, 
    \varepsilon_{2}/\varepsilon_{0}=36.92,
     \lambda\mu^{2}_{\rm d}/\varepsilon_{0}=1.095$ in the case of first
      order phase transition.
This parameter set shows a ratio of spontaneous polarization jump to
 saturated spontaneous polarization about 0.4.
As a comparison with the case of the first order phase transition we
 choose the same parameter set as $\varepsilon_{1}/\varepsilon_{0}=9.48,
  \varepsilon_{2}/\varepsilon_{0}=36.92$ except $\lambda\mu^{2}_{\rm
   d}/\varepsilon_{0}=0$ in the case of second order phase transition.
In the following this set of  parameters is  used in the present
 calculations.

\begin{figure}[btp]
\begin{center}
\includegraphics[width=8cm]{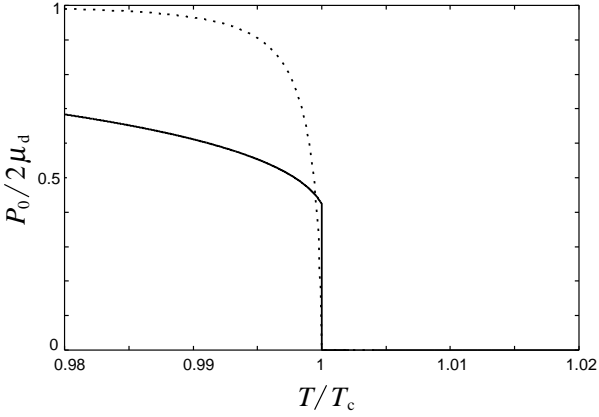}
 \caption{
 The temperature dependence of spontaneous polarization.
 The solid line (first order) and dotted one (second order) are shown.
 In the first order case the spontaneous polarization jump at
 $T=T_{\rm c}$ is about 0.4.
 }
 \label{p0-v1}
\end{center}
\end{figure}

\begin{figure}[tbp]
\begin{center}
\includegraphics[width=8cm]{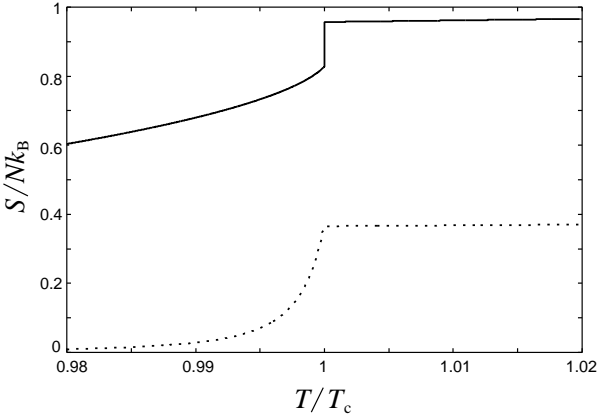}
 \caption{
 The entropy versus temperature.
 The solid line (first order) and dotted one (second order) are shown.
 }
 \label{ent-v1}
\end{center}
\end{figure}

Though without dipole-dipole interaction term  $\lambda=0$ we could
 study the static properties analytically thanks to  explicit expression
  of spontaneous polarization\cite{Ishibashi}, with  $\lambda\ne 0$  we
   cannot obtain analytical spontaneous polarlization anymore.
However, we can deal with  these simultaneous equations numerically by
 making use of the natural iteration method (NIM)\cite{Kiku3} for the
  CVM.
Through the NIM calculation we can obtain the physical quantities such
 as the spontaneous polarization $P_{0}\equiv 2\mu_{\rm d}m_0$ per
  PO$_4$ and the entropy $S$ for each case of the first and the second
   order transition (Fig. \ref{p0-v1} and Fig. \ref{ent-v1}). 
These numerical results will be used to calculate dynamical
 susceptibility.

\begin{figure}[t]
\begin{center}
\includegraphics[width=8cm]{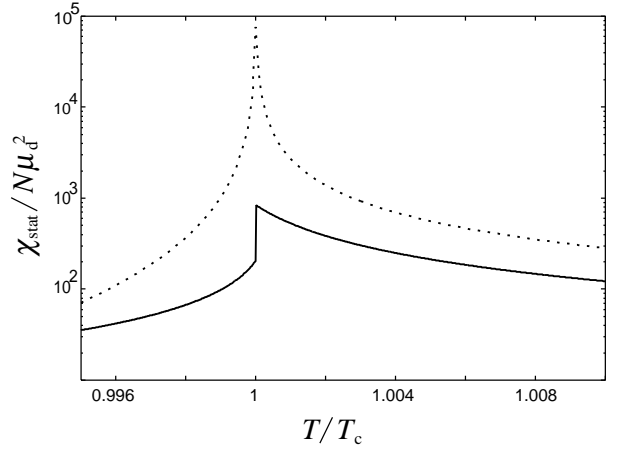}
\caption{
 The uniform susceptibility versus temperature.
 The solid line (first order) and dotted one (second order) are shown.
}
\label{stat-v1}
\end{center}
\end{figure}

The static uniform susceptibility $\chi_{\rm stat}$ of the system  is
 also obtained as the linear response ${\it\Delta} m$ from $m_{0}$
  induced by a uniform field $E$ from the equation of state
   (\ref{polarization}) as
\begin{equation}
\chi_{\rm stat}
\equiv \frac{2N\mu_{\rm d}\mathit{\Delta} m}{E}
= \frac{2N{\mu_{\rm d}}^2}{k_{\rm B}T}
  \frac{1}{X^{-1}-\left(\frac{1}{1-{m_{0}}^2}+2D\right)},
\label{paka}
\end{equation}
with
\begin{equation}
\begin{split}
& A_0=\sqrt{\frac{1+m_0}{1-m_0}} \exp(2 D m_0),\\
& \alpha_0 = 
  \bigl(
   {A_0}^2+{A_0}^{-2}+4\eta_0+4\eta_1(A_0+{A_0}^{-1})+2\eta_2
  \bigr)^{-1},\\
& X=2\alpha_0\bigl({A_0}^2+{A_0}^{-2}+(A_0+{A_0}^{-1})\eta_1\bigr)-2 {m_0}^2,
\end{split}
\end{equation}
where $m_0$ is given from eq.(\ref{polarization}) by
\begin{equation}
m_0=\alpha_0\bigl({A_0}^2-{A_0}^{-2}+2\eta_{1}({A_0}-{A_0}^{-1})\bigr).
\end{equation}
When $D=0$, 
this result reduces to our previous result for the Slater-Takagi model
 without dipole-dipole interaction\cite{Wada}.
The numerical result of uniform susceptibility $\chi_{\rm stat}$ is
 shown in Fig. \ref{stat-v1}.

\section{Dynamical Susceptibility}
By making use of the numerical results obtained in the previous section,
 let us calculate dynamical susceptibility in this section.
When an external electric field $E(t)=E\exp({\rm i}\omega t)$ is
 applied, we assume that
\begin{equation}
m_i(t) = m^0_i+\chi_i(\omega)E\exp({\rm i}\omega t)
\quad (i=1\sim 5)
\label{m5t}
\end{equation}
where the required dynamical susceptibility is given by
 $\chi(\omega)=\mu_{\rm d}\chi_1(\omega)/4$ since it is the response of
  electric polarization to a weak electric field, and $m^{0}_{i}$'s are
   equilibrium order parameters in the absence of an external field.
It should be noted that $m^0_i (i=1\sim 5)$ are obtained
 numerically by the NIM calculation.
By substituting these relations (\ref{m5t}) into (\ref{eqmo}) we get the
 following algebraic equations up to the linear ordre to external field:
\begin{eqnarray}
\left({{\rm i}\omega\tau_{0}}{I}+{M}\right){\mib{\chi}}(\omega)={\mib{b}},\label{mat}
\end{eqnarray}
where $\mib{\chi}(\omega)=(\chi_{1}(\omega)\ \chi_{2}(\omega)\
 \chi_{3}(\omega)\ \chi_{4}(\omega)\ \chi_{5}(\omega))^{t}$
  and  $I$ is the $5\times 5$ unit matrix. 
Since the elements of $5\times5$ matrix $M$ and a $5\times1$ column vector 
 $\mib{b}$ are a little lengthy quantities expressed in terms of only
  $\eta_0, \eta_1, \eta_2$ and $D$, we omit them here.
After numerical calculations by the Gaussian elimination method for a
 fixed $\omega\tau_{0}$ we get the dynamical susceptibility over all the
  temperature region.
With the relaxation times $\tau_{i} (i=1\sim 5)$ and the intensity
 coefficients $\chi^{i}$,
  the dynamical susceptibility per proton $\chi(\omega)$ can be written
   as 
\begin{equation}
\chi(\omega)
 = \frac{\mu_{\rm d}}{4}\chi_{1}(\omega)
 = \sum_{i=1}^{5} \frac{\chi^i}{1+{\rm i}\omega\tau_i}
\label{chiferro},
\end{equation}
where the five relaxation times $\tau_i$ are obtained by a
 diagonalization of $M$ in terms of a matrix $U$ as
\begin{equation}
(UMU^{-1})_{ij}=\frac{\tau_0}{\tau_i}\delta_{ij}\,,\quad
 (\delta_{ij}:\mbox{Kronecker's  delta})
\end{equation}
and the intensity coefficients $\chi^i$ are given by
\begin{equation}
\chi^i
 = \frac{\mu_{\rm d}}{4\tau_0}\tau_i
   \sum^{5}_{j=1}({U}^{-1})_{1i}{U}_{ij}b_j \,.
\end{equation}

Especially in the paraelectric phase we note that, under the inversion of
 an external field $E$, the order parameters $m_1$ and $m_3$ are changed 
into $-m_1$ and $-m_3$, respectively, while $m_2$, $m_4$ and $m_5$ are
 invariant.
Accordingly, $m_1$ and $m_3$ are long range order parameters responding
 linearly to external field $E$, while $m_{2}$, $m_{4}$ and $m_{5}$ are
  short range order parameters responding quadratically to the field.
Thus in the paraelectric phase eq. (\ref{mat}) is reduced to a closed
 algebraic equation in $m_1$ and $m_3$ space and as a result the
  dynamical susceptibility (\ref{chiferro}) can be written down as
\begin{equation}
\chi(\omega)
 = \frac{\chi^1}{1+{\rm i}\omega\tau_{+}}+\frac{\chi^3}{1+{\rm i}\omega\tau_{-}},
\end{equation}
where
\begin{equation}
\begin{split}
& \tau_{\pm}
 = \frac{2}{p\mp\sqrt{p^2-4q}} \quad (\tau_{+}=\tau_1,\tau_{-}=\tau_3)\,,\\
&p = 2\biggl(1+\frac{2\eta_{0}+4\sqrt{\eta_{0}
 \eta_{2}}+2\eta_1(2\sqrt{\eta_0}+\sqrt{\eta_2})}{\Gamma}\\[-5mm]
& \hspace{5cm} -2K^2(1+2D)\biggr)\,,\\
&q = 8K^2(2\sqrt{\eta_{0}}+\sqrt{\eta_{2}})\left(1-\frac{2(1+2D)(1+\eta_1)}{\Gamma}\right)\,,\\
& \chi^{1}
 = \frac{8\beta\mu^2_{\rm d}K^2 \tau_+ \tau_-}{(\tau_+ - \tau_-)\tau_0}
   \left(
    \frac{8(1+\eta_1)(2\sqrt{\eta_0}+\sqrt{\eta_2})}{\Gamma}
    \frac{\tau_+}{\tau_0}-1
   \right)\,,\\
& \chi^{3}
 = \frac{8\beta\mu^2_{\rm d} K^2 \tau_+ \tau_-}{(\tau_+ - \tau_-)\tau_0}
   \left(
    1-\frac{8(1+\eta_1)(2\sqrt{\eta_0}+\sqrt{\eta_2})}\Gamma
    \frac{\tau_-}{\tau_0}
   \right)\,,
\end{split}
\end{equation}
with
\begin{equation}
K = \frac{(1+2\sqrt{\eta_0}+\sqrt{\eta_2})\sqrt{\eta_1}}\Gamma\,.
\end{equation}

\begin{figure}[tbp]
\begin{center}
\includegraphics[width=8cm]{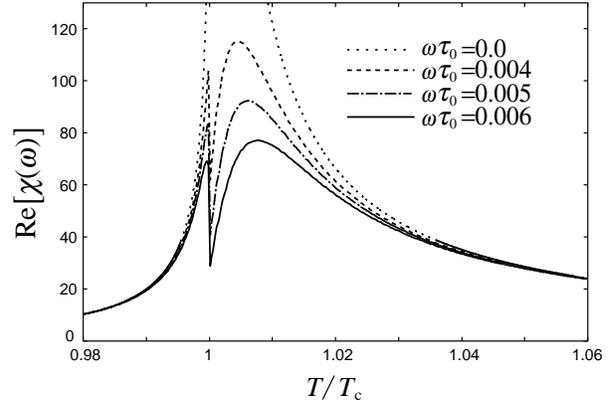}
\caption{
The real part of dynamical susceptibility in the case of first order
 phase transition.
}
\label{dyreal2}
\end{center}
\end{figure}
\begin{figure}[tbp]
\begin{center}
\includegraphics[width=8cm]{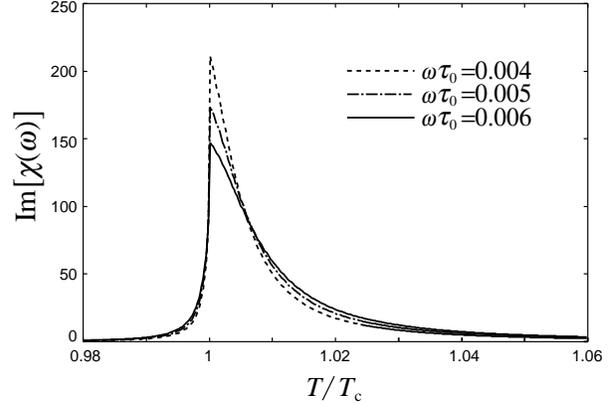}
\caption{
The imaginary part of dynamical susceptibility in the case of first
 order phase transition.
}
\label{dyimagl750}
\end{center}
\end{figure}
\begin{figure}[tbp]
\begin{center}
\includegraphics[width=8cm]{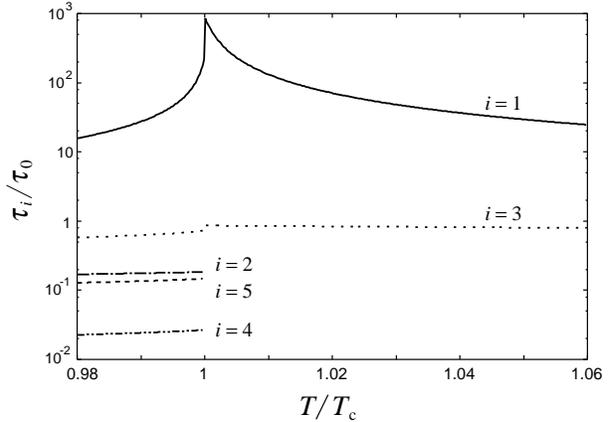}
\caption{
The five relaxation times versus temperature in the case of first order
 phase transition.
In the paraelectric phase $\tau_2,\,\tau_4,\,\tau_5$ are omitted
 because any contribution to $\chi(\omega)$ from them do not exist.
}
\label{relax2}
\end{center}
\end{figure}
\begin{figure}[tbp]
\begin{center}
\includegraphics[width=8cm]{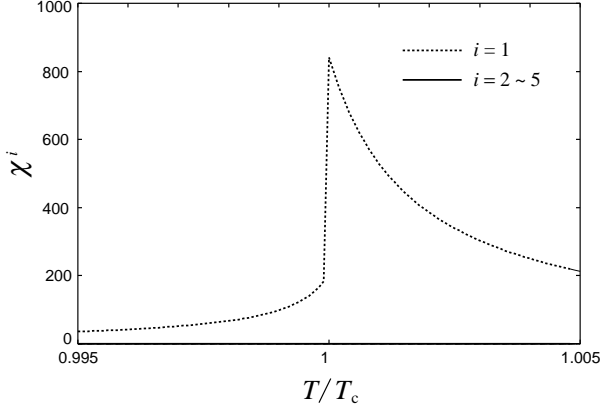} \caption{ The five
intensities versus temperature in the case of first order phase
transition.  One of the intensities $\chi^1$ becomes dominant and
$\chi^2 \sim \chi^5$ have almost zero value in comparison with $\chi^1$
over all temperature region.  } \label{intl750}
\end{center}
\end{figure}

Next, we plot the real and imaginary part of dynamical susceptibility
 over all the temperature region in the first order transition in
  Fig. \ref{dyreal2} and Fig. \ref{dyimagl750}.
We see a small dip in the real part of
 $\chi(\omega)$(Fig. \ref{dyreal2}) consistent with the
  experiments\cite{Koz}.
We notice from Fig. \ref{relax2} and Fig.  \ref{intl750} that one of the
 five modes overwhelms other four modes in relaxation time $\tau_i$ and
  intensity coefficient $\chi^i$,  though any relaxation time
   $\tau_i$ is finite at $T=T_{\rm c}$ in the first order phase
    transition. 
That is the reason why the small dip structure appears around the
 transition temperature in the real part of the dynamical
  susceptibility. 
On the other hand, as shown in I, in the Slater-Takagi model the phase
 transition always becomes the second order and the most dominant mode
  shows a critical slowing down which leads to the vanishing of the real
   part of dynamical susceptibility at the transition temperature.
The imaginary part of dynamical susceptibility is also in qualitative
 agreement with the experiments\cite{Koz}.

\section{Light Scattering Intensity}
\begin{figure}[tbp]
\begin{center}
\includegraphics[width=8cm]{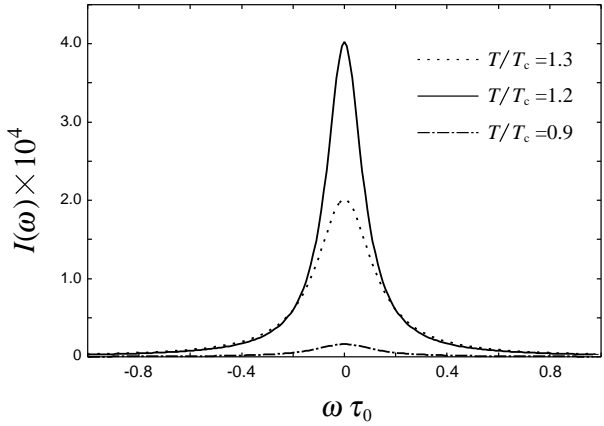}
\caption{
The frequency dependence of light scattering intensity in the case of
 first order phase transition.
}
\label{int1}
\end{center}
\end{figure}

Let us calculate the light scattering intensity assuming the first order
 transition with the same previous set of parameters and compare it with
  the experimental results.
The light scattering intensity $I(\omega)$ is expressed in terms of the
 imaginary part of $\chi(\omega)$ in the high temperature approximation
  as
\begin{eqnarray}
I(\omega) \propto \frac{{\rm Im}[\chi(\omega)]}{\omega}.
\end{eqnarray}
The numerical results of $I(\omega)$ are shown in Fig. \ref{int1}. 
When the transition temperature is approached from high temperature
 side above $T_{\rm c}$, the height of a peak increases and the width of
 the peak decreases around $\omega =0$\cite{Matu}.
However, since it is assumed to be the first order transition, the
 height of the peak becomes the largest at the transition temperature
  but still finite.
Moreover, as the temperature is decreased across $T_{\rm c}$, the height
 of the peak around $\omega=0$ is decreased drastically. 
The central mode at $\omega=0$ is  caused by the dipolar fluctuations of
 relaxation modes in the present model.
Further, since from the imaginary part of eq.(\ref{chiferro}) it is easily
 obtained as
\begin{eqnarray}
\int_{-\infty}^{\infty} I(\omega ){\rm d}\omega
 \propto \int_{-\infty}^{\infty} {\rm d}\omega
\sum_{i}\frac{\chi^i\tau_i}{1+(\omega\tau_i)^2}
 = \pi \sum_{i=1}^5 \chi^i
\label{intchi}
\end{eqnarray}
these behaviors are easily understood as follows.
Since the total integral value of scattering intensity over $\omega$ is
 proportional to the sum of intensity of relaxation modes at each
  temperature $T$, the largest intensity over $\omega$ is reached at the
   transition temperature $T_{\rm c}$ but does not diverge in the present
    first order transition.
These behaviors agree qualitatively with the experimental data.
\cite{Tominaga}

\section{Conclusions and Discussions}
In order to explain a small dip structure around the ferroelectric
 transition  temperature in the real part of dynamical susceptibility of
  KDP crystal observed in the experiment, we extend our previous
  calculation in the Slater-Takagi model to the Slater-Takagi model with 
   phenomenologically introduced dipole-dipole interaction.
We applied the path probability method(PPM) to the Slater-Takagi model
 with dipole-dipole interaction in the tetragonal cactus approximation
  which can take care of the ice rule characteristic of KDP crystal.
With the dipole-dipole interaction being increased, the phase transition
 changes from the second order into the first order transition.
Further, in the present model the phase transition shows the
 order-disorder transition of proton configurations.
We see the dynamical fluctuations of electric polarization modes due to
 proton configuration in the dynamical susceptibility which is expressed
  in terms of relaxation times of the electric polarizations with
   intensity of each mode.
The Slater-Takagi model always induces the second order transition and
 shows a critical slowing down as to relaxation time of the relevant
  mode.
Then the system can not follow the external electric field with finite
 frequency at the transition temperature $T_{\rm c}$ due to the critical
  slowing down.
This is the reason why the real part of the dynamical susceptibility
 gives zero at the critical temperature $T_{\rm c}$.
On the other hand, the Slater-Takagi model with the appropriate
 dipole-dipole interaction undergoes the first order transition and the
  relaxation time of each mode becomes finite.
Thus the system can manage to follow the frequency dependent external
 field, though the relaxation time of relevant mode  is relatively
  long because of the first order transition close to the second order
   transition of KDP.
This fact explains why the dip structure is observed around the
 transition temperature in the experiment of the real part of dynamical
  susceptibility.
We also calculated the light scattering intensity utilizing the
 imaginary part of the present dynamical susceptibility.
Since our model deals with relaxation modes in the order-disorder
 transition, we obtain only the central peak structure around $\omega=0$.
This peak structure increases toward the transition temperature, though
 the height of the peak does not diverges even the transition
  temperature in the present first order transition.
This behavior is in good agreement with the data of Tominaga {\it et
 al.}\cite{Tominaga}.
 
Finally we should note that all our calculations show not quantative but 
 qualitatively good agreements with experiments which are ascribed to
  the Slater-Takagi model with phenomenologically introduced
   dipole-dipole interaction.

\end{document}